\documentclass[a4paper,11pt]{article}
\usepackage{pos}

\usepackage{amsmath,bm}
\graphicspath{{figures/}}
\usepackage[cache=true,newfloat=true,frozencache]{minted}
\usepackage[capitalize]{cleveref}

\setminted[json]{fontsize=\scriptsize, numbersep=5pt, frame=lines, framesep=2mm, gobble=0}
\setminted[python]{fontsize=\scriptsize, numbersep=5pt, frame=lines, framesep=2mm, gobble=0, linenos}
\newcommand{\HiFa}{\texttt{HistFactory}}

\newcommand{\pyhf}{\texttt{pyhf}}

\newcommand{\term}[1]{\textsl{#1}}

\newcommand{\freeset}{\bm{\eta}}
\newcommand{\constrset}{\bm{\chi}}
\newcommand{\singleconstr}{\chi}

\newcommand{\channelcounts}{\bm{n}}
\newcommand{\auxdata}{\bm{a}}

\newcommand{\poiset}{\bm{\psi}}
\newcommand{\nuisset}{\bm{\theta}}

\newcommand{\fullset}{\bm{\phi}}

\newcommand{\annoterel}[2]{%
 \overset{%
  \substack{\hidewidth\text{#1}\hidewidth\\\downarrow}%
 }{#2}%
}

\newcommand{\annotereldn}[2]{%
 \underset{%
  \substack{\uparrow\\\hidewidth\text{#1}\hidewidth}%
 }{#2}%
}

\newcommand*{\TeV}{\ensuremath{\text{Te\kern -0.1em V}}}
\newcommand*{\GeV}{\ensuremath{\text{Ge\kern -0.1em V}}}
\newcommand*{\MeV}{\ensuremath{\text{Me\kern -0.1em V}}}
\newcommand*{\keV}{\ensuremath{\text{ke\kern -0.1em V}}}
\newcommand*{\eV}{\ensuremath{\text{e\kern -0.1em V}}}

\title{pyhf: pure-Python implementation of HistFactory with tensors and automatic differentiation}
\ShortTitle{pyhf: pure-Python implementation of HistFactory}

\author*[a]{Matthew Feickert}
\author[b]{Lukas Heinrich}
\author[c]{Giordon Stark}

\affiliation[a]{University of Wisconsin-Madison,\\
447 Lorch St., Madison, WI, USA}

\affiliation[b]{Technical University Munich,\\
Arcisstraße 21, 80333 München, Germany}

\affiliation[c]{University of California Santa Cruz SCIPP,\\
Santa Cruz, CA, USA}

\emailAdd{matthew.feickert@cern.ch}
\emailAdd{lukas.heinrich@cern.ch}
\emailAdd{giordon.holtsberg.stark@cern.ch}

\abstract{%
 The HistFactory p.d.f. template is per-se independent of its implementation in ROOT and it is useful to be able to run statistical analysis outside of the ROOT, RooFit, RooStats framework.
\pyhf{} is a pure-Python implementation of that statistical model for multi-bin histogram-based analysis and its interval estimation is based on the asymptotic formulas of ``Asymptotic formulae for likelihood-based tests of new physics''.
\pyhf{} supports modern computational graph libraries such as TensorFlow, PyTorch, and JAX in order to make use of features such as auto-differentiation and GPU acceleration.
In addition, \pyhf{}'s JSON serialization specification for HistFactory models has been used to publish 23 full probability models from published ATLAS collaboration analyses to HEPData.

}

\FullConference{%
  41st International Conference on High Energy physics - ICHEP2022\\
  6-13 July, 2022\\
  Bologna, Italy
}

\begin{document}
\maketitle
\section{Introduction}\label{sec:introduction}
Measurements in High Energy Physics (HEP) aim to determine the compatibility of observed events with theoretical predictions.
The relationship between them is often formalised in a statistical \term{model} $f(\bm{x}|\fullset)$ describing the probability of data $\bm{x}$ given model parameters $\fullset$.
Given observed data, the \term{likelihood} $\mathcal{L}(\fullset)$ then serves as the basis to test hypotheses on the parameters~$\fullset$.
For measurements based on binned data (\term{histograms}), the \HiFa{}~\cite{Cranmer:1456844} family of statistical models has been widely used for likelihood construction in both Standard Model (SM) measurements (e.g. Refs.~\cite{HIGG-2013-02,Aaij:2015sqa}) as well as searches for new physics (e.g. Ref.~\cite{ATLAS-CONF-2018-041}) and reinterpretation studies (e.g. Ref.~\cite{Heinrich:2018nip}).
\pyhf{}~\cite{pyhf,pyhf_joss} is presented as the first pure-Python implementation of the \HiFa{} specification.
In addition to providing a Python and command line API for \HiFa{} model building and inspection, it leverages modern open source $n$-dimensional array libraries to take advantage of automatic differentiation and hardware acceleration to accelerate the statistical inference and reduce the time to analyst insight.

\section{\HiFa{} Formalism}\label{sec:HistFactory}

\HiFa{} statistical models --- described in depth in Ref.~\cite{ATL-PHYS-PUB-2019-029} and Ref.~\cite{CHEP_2019}  --- center around the simultaneous measurement of disjoint binned distributions (\term{channels}) observed as event counts $\channelcounts$.
For each channel, the overall expected event rate is the sum over a number of physics processes (\term{samples}).
The sample rates may be subject to parametrised variations, both to express the effect of \term{free parameters} $\freeset$ and to account for systematic uncertainties as a function of \term{constrained parameters} $\constrset$, whose impact on the expected event rates from the nominal rates is limited by \term{constraint terms}.
In a frequentist framework these constraint terms can be viewed as \term{auxiliary measurements} with additional global observable data $\auxdata$, which paired with the channel data $\channelcounts$ completes the observation $\bm{x} = (\channelcounts,\auxdata)$.
The full parameter set can be partitioned into free and constrained parameters $\fullset = (\freeset,\constrset)$, where a subset of the free parameters are declared \term{parameters of interest} (POI) $\poiset$ (e.g. the \term{signal strength}) and all remaining parameters as \term{nuisance parameters} $\nuisset$.

\begin{equation}
 \label{eqn:parameters_partitions}
 f(\bm{x}|\fullset) = f(\bm{x}|\annoterel{free}{\freeset},\annotereldn{constrained\hspace{1cm}}{\constrset}) = f(\bm{x}|\annoterel{\hspace{2cm}parameters of interest}{\poiset},\annotereldn{\hspace{1cm}nuisance parameters}{\nuisset})
\end{equation}

The overall structure of a \HiFa{} probability model is then a product of the {\color{blue}analysis-specific model term} describing the measurements of the channels and the {\color{red}analysis-independent set of constraint terms}:
\begin{equation}
 \label{eqn:hifa_template}
 f(\channelcounts, \auxdata \,|\,\freeset,\constrset) = \underbrace{\color{blue}{\prod_{c\in\mathrm{\,channels}} \prod_{b \in \mathrm{\,bins}_c}\textrm{Pois}\left(n_{cb} \,\middle|\, \nu_{cb}\left(\freeset,\constrset\right)\right)}}_{\substack{\text{Simultaneous measurement}\\%
   \text{of multiple channels}}} \underbrace{\color{red}{\prod_{\singleconstr \in \constrset} c_{\singleconstr}(a_{\singleconstr} |\, \singleconstr)}}_{\substack{\text{constraint terms}\\%
   \text{for ``auxiliary measurements''}}},
\end{equation}
where within a certain integrated luminosity one observes $n_{cb}$ events given the expected rate of events $\nu_{cb}(\freeset,\constrset)$ as a function of unconstrained parameters $\freeset$ and constrained parameters $\constrset$.
The latter has corresponding one-dimensional constraint terms $c_\singleconstr(a_\singleconstr|\,\singleconstr)$ with auxiliary data $a_\singleconstr$ constraining the parameter $\singleconstr$.
The expected event rates $\nu_{cb}$ are defined as

\begin{equation}
 \label{eqn:sample_rates}
 \nu_{cb}\left(\fullset\right) = \sum_{s\in\mathrm{\,samples}} \nu_{scb}\left(\freeset,\constrset\right) = \sum_{s\in\mathrm{\,samples}}\underbrace{\left(\prod_{\kappa\in\,\bm{\kappa}} \kappa_{scb}\left(\freeset,\constrset\right)\right)}_{\text{multiplicative modifiers}}\, \Bigg(\nu_{scb}^0\left(\freeset, \constrset\right) + \underbrace{\sum_{\Delta\in\bm{\Delta}} \Delta_{scb}\left(\freeset,\constrset\right)}_{\text{additive modifiers}}\Bigg)
\end{equation}

\noindent from constant \term{nominal rate} $\nu_{scb}^0$ and a set of multiplicative and additive \term{rate modifiers} $\bm{\kappa}(\fullset)$ and $\bm{\Delta}(\fullset)$.

\section{\pyhf{}}\label{sec:pyhf}

Through adoption of open source $n$-dimensional array (``tensor'' in the machine learning world) computational Python libraries, pyhf{} decreases the abstractions between a physicist performing an analysis and the statistical modeling without sacrificing computational speed.
By taking advantage of tensor calculations and hardware acceleration, pyhf{} can achieve comparable or better performance than the \texttt{C++} implementation of \HiFa{} on data from real LHC analyses in most situations.
pyhf{}'s default computational backend is built from NumPy and SciPy, and supports TensorFlow, PyTorch, and JAX as alternative backend choices.
These alternative backends support hardware acceleration on GPUs, and in the case of JAX JIT compilation, as well as auto-differentiation allowing for calculating the full gradient of the likelihood function --- all contributing to speeding up fits.

\subsection{JSON Schema}\label{subsec:HistFactory_schema}

The structure of the JSON specification of \HiFa{} models~\cite{ATL-PHYS-PUB-2019-029} used by \pyhf{} closely follows the original XML-based specification~\cite{Cranmer:1456844}.
The JSON specification for a \HiFa{} \term{workspace} is a primary focus of Ref.~\cite{ATL-PHYS-PUB-2019-029}, but a workspace can be summarised as consisting of a set of channels (an analysis region) that include samples and possible parameterised modifiers, a set of measurements (including the POI), and observations (the observed data).
\Cref{lst:example:2binchannel} demonstrates a simple workspace representing the measurement of a single two-bin channel with two samples: a signal sample and a background sample.
The signal sample has an unconstrained normalisation factor $\mu$, while the background sample carries an uncorrelated shape systematic.
The background uncertainties for the bins are 10\% and 20\% respectively.
Use of this JSON specification has allowed for the publication of 23 full statistical models from ATLAS analyses to HEPData at the time of writing in 2022.
This has been a significant step forward in enabling reinterpretation and recasting of LHC results by the broader particle physics community~\cite{Cranmer:2021urp}.

\begin{listing}
\begin{minted}{json}
{
    "channels": [
        { "name": "singlechannel",
          "samples": [
            { "name": "signal",
              "data": [5.0, 10.0],
              "modifiers": [ { "name": "mu", "type": "normfactor", "data": null} ]
            },
            { "name": "background",
              "data": [50.0, 60.0],
              "modifiers": [ {"name": "uncorr_bkguncrt", "type": "shapesys", "data": [5.0,12.0]} ]
            }
          ]
        }
    ],
    "observations": [
        {"name": "singlechannel", "data": [50, 60]}
    ],
    "measurements": [
        { "name": "Measurement", "config": {"poi": "mu", "parameters": []} }
    ]
}
\end{minted}
 \caption{A toy example of a 2-bin single channel workspace with two samples.
  The signal sample has expected event rates of 5.0 and 10.0 in each bin, while the background sample has expected event rates of 50.0 and 60.0 in each bin.
  An experiment provided the observed event rates of 50.0 and 60.0 for the bins in that channel.
  The uncorrelated shape systematic on the background has 10\% and 20\% uncertainties in each bin, specified as absolute uncertainties on the background sample rates.
  A single measurement is defined which specifies $\mu$ as the POI~\cite{ATL-PHYS-PUB-2019-029}.}
 \label{lst:example:2binchannel}
\end{listing}

\subsection{Enabling Analysis Ecosystems}\label{subsec:analysis_ecosystems}

\begin{figure}
    \centering
    \includegraphics[width=0.8\linewidth]{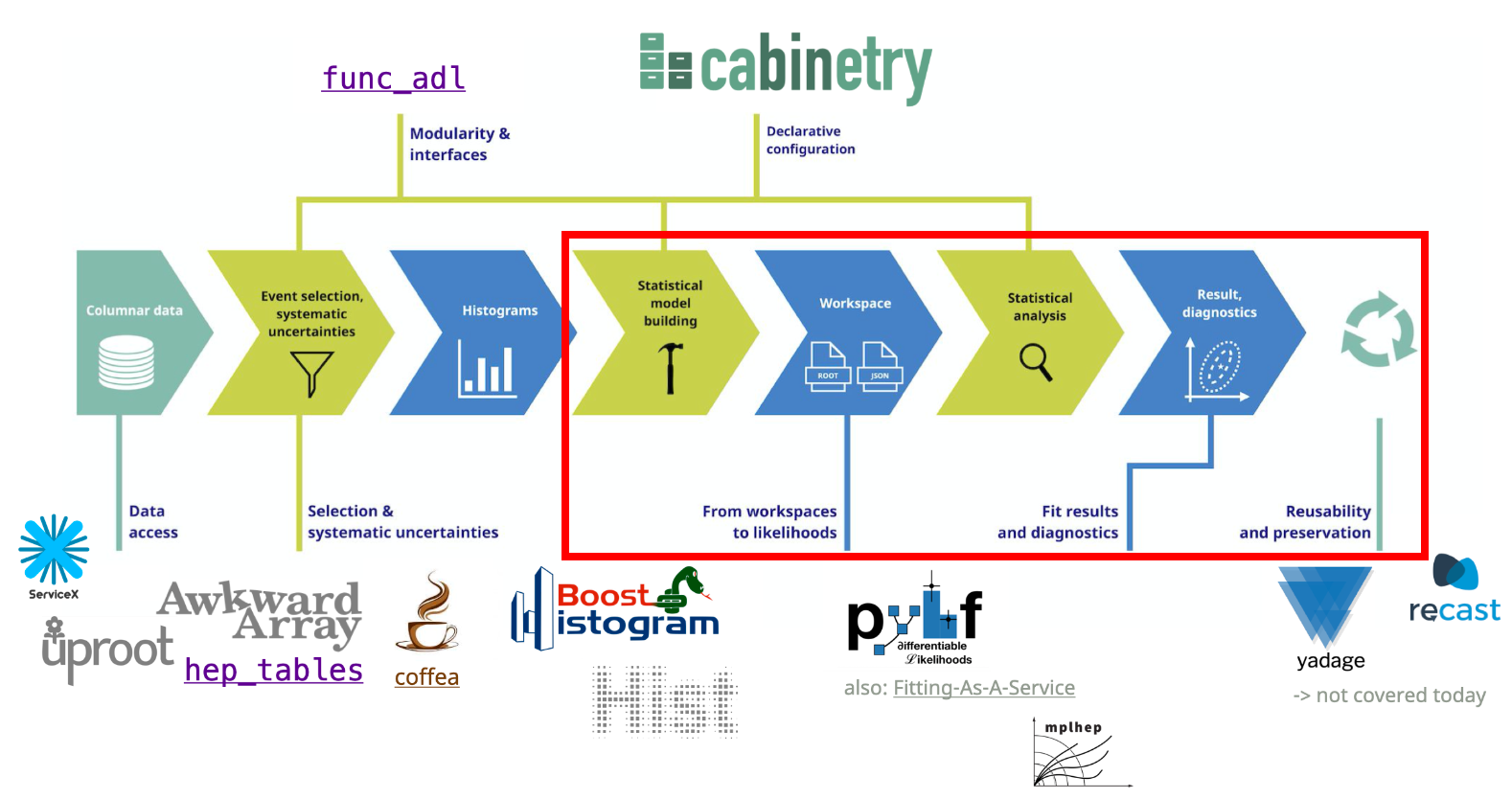}
    \caption{Overview of the IRIS-HEP Analysis Systems pipeline for analysis in the HL-LHC era.
The red outline indicates the areas of the pipeline in which \pyhf{} is used either directly or as an underlying library.}
    \label{fig:analysis-systems-pipeline}
\end{figure}

In addition to being used in ATLAS analyses, and in the flavor physics community~\cite{Belle-II:2021rof,Belle:2022gbl}, \pyhf{} has been used as a computational engine for reinterpretation studies by the particle physics phenomenology community~\cite{Alguero:2020grj,Alguero:2020yhu} and as the inference engine for Scikit-HEP library \texttt{cabinetry}~\cite{cabinetry}, as well as other more analysis specific open source projects~\cite{abcd_pyhf,Simpson:2022suz}.
The adoption of \pyhf{} as a library for other projects to build upon has large implications for establishing standards and providing improvements across ecosystems of analysis tools.
Of particular note, the Institute for Research and Innovation in Software for High Energy Physics (IRIS-HEP)~\cite{iris-hep} has adopted \pyhf{} as a core part of its Analysis Systems pipeline --- a demonstrator model for modern distributed computing for experiments in the high-luminosity LHC (HL-LHC) era --- which has provided rigorous testing of its interoperability with other tools.
Improvements to \pyhf{} directly impact all the areas highlighted in~\Cref{fig:analysis-systems-pipeline}.
In addition to its computational abilities, \pyhf{} is highly portable given its pure-Python nature and use of dependencies, like SciPy, that are broadly trusted in computational science and have been built for ubiquitous architectures.
This allows for full \pyhf{} runtimes to be natively used in novel environments, such as the Pyodide port of CPython to WebAssembly/Emscripten.
While Pyodide is not optimal for serious computational use cases, the ability to use the full \pyhf{} API allows for creation of statistical linting and visualization tools that use the same tooling as in production while leveraging interactivity of web native platforms enabled by Pyodide and the PyScript framework.

\section{Conclusions}\label{sec:conclusions}

\pyhf{} is the first pure-Python implementation of the \HiFa{} specification that leverages modern open source $n$-dimensional array libraries as computational backends to exploit automatic differentiation and hardware acceleration to speed up fits and reduce the time to scientific insight.
It provides a Python and command line API for building, inspection, and to perform statistical inference for \HiFa{} models, and its JSON model serialization has enabled publication of full statistical models from the ATLAS collaboration and improved reinterpretations.
As \pyhf{} is an open source library that has been built as part of the Scikit-HEP community project it has been readily adopted by a growing number of other libraries and tools as a computational and inference engine, allowing for improvements in the library API and computational backends to propagate to the broader user community.
Growing community support and interaction, adoption across the broader particle physics community, and rigorous testing from LHC experiments and IRIS-HEP systems has demonstrated that \pyhf{} has become a key component of the growing ecosystem of Pythonic open source scientific tools in particle physics.

\acknowledgments{%
Matthew Feickert's contributions to this work were supported by the U.S. National Science Foundation (NSF) under Cooperative Agreement OAC-1836650 (IRIS-HEP).
Lukas Heinrich is supported by the Excellence Cluster ORIGINS, which is funded by the Deutsche Forschungsgemeinschaft (DFG, German Research Foundation) under Germany's Excellence Strategy -- EXC-2094-390783311.
}

\clearpage
\bibliographystyle{JHEP}
\bibliography{bib/ref,bib/ATLAS,bib/ConfNotes,bib/PubNotes,bib/reinterpretation}

\end{document}